\documentclass[%
 aip,
 long, numerical bibliography, (default)
 jmp,%
 amsmath,amssymb,
reprint,%
]{revtex4-1}

\usepackage{graphicx}
\usepackage{dcolumn}
\usepackage{bm}
\begin{document}

\preprint{AIP/123-QED}

\title{Facile fabrication of lateral nanowire wrap-gate devices with improved performance}
\author{Sajal Dhara}
\homepage{dhara@tifr.res.in}
\author{Shamashis Sengupta}
\author{Hari S. Solanki}
\author{ Arvind Maurya}
\author{ Arvind Pavan R.}
\author{M. R. Gokhale}
\author{Arnab Bhattacharya}
\author{Mandar M. Deshmukh}
 \homepage{deshmukh@tifr.res.in}
\affiliation{Department of Condensed Matter Physics and Materials Science, Tata Institute of Fundamental Research, Homi Bhabha Road, Mumbai 400005 India}
\date{\today}

\begin{abstract}
We present a simple fabrication technique for lateral nanowire wrap-gate devices with high capacitive coupling and field-effect
mobility. Our process uses e-beam lithography with a single resist-spinning step, and does not require chemical etching. We measure, in the temperature range 1.5-250~K, a subthreshold slope of 5-54~mV/decade and mobility of 2800-2500~$cm^2/Vs$ -- significantly larger than previously reported lateral wrap-gate devices. At depletion, the barrier height due to the gated region is proportional to applied wrap-gate voltage.
\end{abstract}

\maketitle

Nanowire field effect transistors (NWFETs) have shown promise for device applications in the field of nanoprocessor and sensors\cite{application,biosense1,Horizontal_sensor,vertical_sensor}.
Several strategies for improving NWFET device performance have been reported\cite{samu1,samu2,storm1}. A major challenge is to increase the capacitive coupling between the gate and the nanowire (NW) for a better control of the gate response for high frequency applications\cite{egard}. A wrap-gate NWFET, that has a coaxial gate electrode around the NW device, is an ideal device geometry where the gate electrode can control the charge transport effectively. Previous reports on vertical wrap-around gate NWFET devices made on nanowire arrays have shown improved performance. However, the fabrication typically involves several lithography steps\cite{samu1,samu2}. In a recent work, Storm \emph{et al.}\cite{storm2} have successfully fabricated a lateral wrap-gate; such devices demonstrate the potential of wrap-gate NWFET, however; the fabrication process involves several steps of chemical-etching following deposition of oxides and metal onto the as-grown NWs (possible only for NWs which are vertically grown and of low density). It is desirable to develop a generalized method of fabrication that eliminates wet etching and multiple lithographic steps to avoid damage to the surface of the NWs which can result in poor mobility.

In this letter, we report a simple fabrication process for lateral wrap-gate NWFET devices with a single step of resist spinning together with e-beam lithography to define source, drain and gate electrodes without any etching steps.
We demonstrate n-type FET behavior with the application of wrap-gate voltage (V$_g$) with a large current on-off ratio ($5\times 10^3$). In the depletion region, we show how the activation energy, which is a measure of the potential barrier in the gated region with respect to the un-gated parts of the nanowire, varies with the application of V$_g$. We characterize the capacitance of these wrap-gate NWFETs, as well as its subthreshold slope (SS). The temperature variation (1.5-250K) of SS (5-54mV/decade) shows how the performance of these devices improve at low temperatures. The estimated mobility (2800-2500 $cm^2/Vs$ over the aforementioned temperature range) is an order of magnitude higher than the recently reported mobility of a wrap-gate device \cite{storm2}.
\begin{figure}
\includegraphics[width=70mm, bb=0 0 397 277]{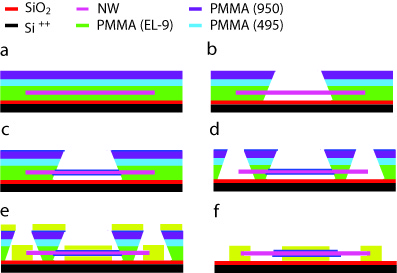}
\caption{ Various steps of lithography used for fabricating wrap-gate devices. a) Sandwiched nanowire between polymer e-beam resists of different molecular weights. b) The gate electrode is patterned by e-beam and developed using a standard developer. c) 10~nm thick HfO$_2$ is conformally coated by ALD. d) The source and drain electrodes are written and developed leading to the structure where the NW is suspended in three segments. e) The deposition of the electrodes using chromium and gold by DC magnetron sputtering leads to fabrication of drain, source and gate electrodes. f) The last step consists of liftoff in acetone to remove the polymer resist and metal layers.}
\end{figure}

InAs nanowires, $\sim10-15\mu$m long and $\sim$70-100nm in diameter, are grown via the vapour-liquid-solid technique as described previously \cite{sajal_prb}. As-grown nanowires are sonicated in isopropanol (IPA) and dispersed on a predefined marker patterned on a 300~nm SiO$_2$ substrate coated with PMMA (Microchem EL9). These NWs are then sequentially covered with another 3 layers of PMMA (EL-9, PMMA495, PMMA950). Fig. 1a shows the schematic of NW sandwiched between layers of e-beam resist. First, an e-beam exposure is done to define the wrap-gate electrode; this stage allows the length of the wrap-gate, along the length of the NW, to be precisely defined. After developing this exposure in a mixture of methyl isobutyl ketone (MIBK) and IPA, the NW is suspended in the developed region that defines the length of the wrap-gate -- as shown in Fig. 1b. A gate dielectric consisting of $\sim$10 nm of HfO$_2$ is coated conformally around the nanowire by atomic layer deposition (ALD) at a temperature of 120$^o$C (Fig. 1c). It is critical to use a low-temperature process for HfO$_2$ deposition to prevent hardbaking the e-beam resist layers (ALD process is given in supplementary material\cite{supp}). Following this, we pattern the source and drain electrodes by another e-beam exposure, through the thin HfO$_2$ layer coated all over the resist. The electron beam at 20~kV can expose the underlying polymer resist layers and this is the key reason that our process reduces several steps of fabrication. Another step of e-beam development in MIBK:IPA leads to the formation of patterns for the source and drain electrodes (shown in Fig. 1d). Before depositing the electrodes we use a NH$_4$S$_x$ treatment for getting better Ohmic contacts (described elsewhere \cite{passivation}) at the source and drain electrodes and find that this does not compromise the gate dielectric. Deposition of 100~nm Cr and 300~nm Au in a DC sputtering system defines the source, drain and gate electrodes in a single metallization step (a schematic at the end of this step is shown in Fig. 1e). Sputter deposition of metal is critical as the metal clusters have broader momentum distribution and this ensures that the metal is deposited under the NWs leading to a conformal deposition of metal all around the NW at the electrodes. The final step consists of a lift-off in acetone to get the wrap-around gate device (schematic of the device shown in Fig. 1f). The liftoff step crucially depends on the stack of multilayered resist used in sandwiching the NW as described earlier. Fig. 2a shows a scanning electron microscope image of a finished wrap-gate NWFET, also shows the suspended segments between the wrap-gate and other two electrodes (device images with different wrap-gate lengths are shown in supplementary material\cite{supp}). The yield of our technique for wrap-gate device is 100 percent. This technique does not require chemical etching and thus applicable not only for InAs but can be used for NWs of other materials as well.
\begin{figure}
\includegraphics[width=85mm, bb=0 0 552 418.5]{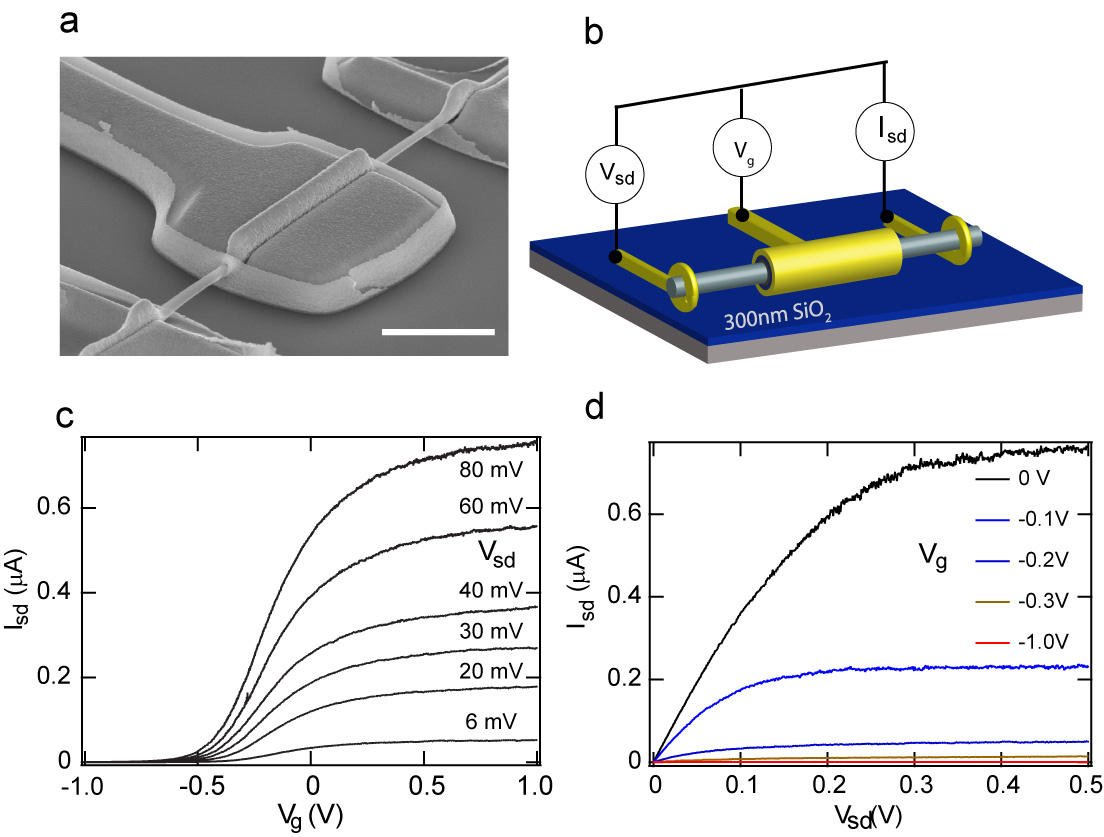}
\caption{ (a) A tilted SEM image of a device. The scale bar corresponds to 1~$\mu$m. The NW is suspended 200~nm above the SiO$_2$ substrate. (b) Schematic of the wrap-gate nanowire device and the circuit used for electrical transport. (c) At room temperature, I$_{SD}$ as a function of V$_g$ at different applied V$_{SD}$. (d) Room temperature I-V characteristic at different applied V$_g$, shows good saturation characteristics.
}
\end{figure}
Fig. 2a shows a tilted angle SEM of a NWFET and the schematic of the device together with the circuit used for measurement (Fig. 2b). The room temperature response of the source-drain current (I$_{SD}$) with V$_g$ for different applied source-drain voltage (V$_{SD}$) is shown in Fig. 2c, and demonstrates low voltage operation of a typical NWFET.
 The I-V shown in Fig. 2d, shows good saturation characteristics, the current on-off ratio being $\sim 5\times 10^3$. We have studied the temperature dependence of the conductance as a function of V$_g$ down to 1.5K. Fig.~3a shows conductance as a function of V$_g$. The temperature variation of SS (at zero V$_{SD}$) can be seen in the inset of Fig. 3a -- a value of 54mV/decade at 250K reduces to ~5mV/decade at 1.5K. The conductance is thermally activated near the off state of the FET (see Fig. 3b) due to the potential barrier formed under the wrap-gate (see Fig. 3c). The activation energy decreases linearly with applied V$_g$ (Fig. 3c); maximum value of $\sim$0.17 eV occurs when the device is completely turned off. This shows that the gate can effectively push the Fermi level near the center of the band gap (0.35 eV). The wrap-gate covers the middle part of the NW ($\sim$ 4 $\mu$m) and this length can be varied in the lithography step (Fig. 1b). By varying the wrap-gate potential, we are changing the conductance G$_2$ of the active region while the two neighboring segments of the nanowire having conductance G$_1$ and G$_3$ are independent of V$_g$. This activation behavior can be understood from the schematic of the band diagram of the wrap-gated NW given in the inset of Fig. 3c, together with the resistor model, that is used later to extract field effect mobility.
\begin{figure}
\includegraphics[width=85mm, bb=0 0 1059 821]{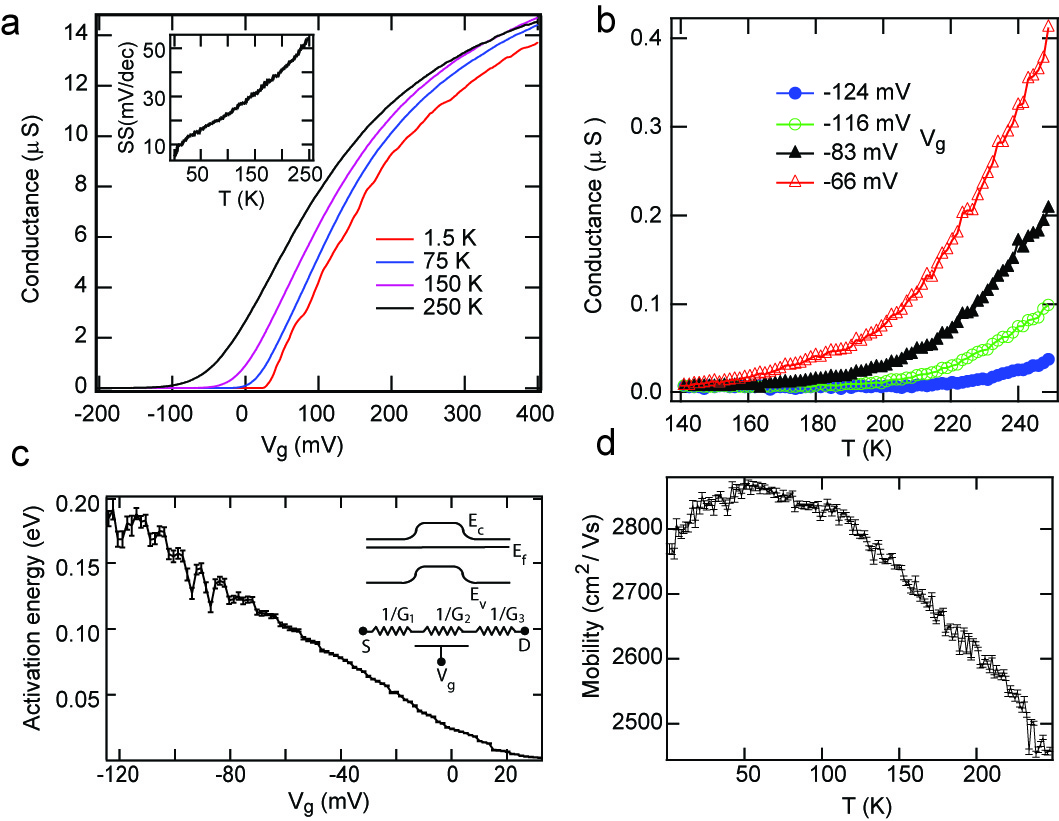}
\caption{ \label{fig:figure2}(a) Field effect transistor (FET) behavior of the device at different temperatures. Inset shows the plot of the subthreshold slope vs temperature. (b) Conductance as a function of temperature when the nanowire is depleted. (c) Shows the variation of the height of the potential barrier as a function of V$_g$. Inset shows the cartoon of the band diagram of the device. (d) Plot of field effect mobility as a function of temperature, shows that mobility peaks around 50K.}
\end{figure}
One important parameter of a FET is its capacitive coupling with the gate electrode. We have done a capacitance-voltage (C-V) measurement at room temperature to get an accurate value of capacitance of the dielectric between the wrap-gate electrode and the nanowire. The measured capacitance per unit length is around $\sim1.5fF/\mu m$ (details of the measurement and frequency response is given in supplementary material\cite{supp}), much higher than the $\sim 100 aF/\mu m$ range typically seen in
cylinder-on-plane device configurations\cite{ford}.

Another parameter to benchmark a FET is its field effect mobility. We have estimated field effect mobility($\mu$) from the
slope of conductance(G) vs V$_g$ plot ($\frac{dG}{dV_g}$), using the expression $\mu=\frac{L^2}{C}\frac{dG}{dV_g}$, where L and C are the length of the nanowire and capacitance of the gate dielectric respectively. The $\frac{dG}{dV_g}$, only due to the active part of the NWFET, is estimated assuming that at some V$_g$ near flat band, the wire will conduct homogeneously in all the three parts
(this assumption gives us only a lower bound to the mobility; details are given in supplementary material\cite{supp}).

Fig. 3d shows the field effect mobility value 2500-2800 cm$^2/Vs$, this is comparable to the value obtained previously on standard InAs NWFET of cylinder-on-plane geometry \cite{sajal_prb,thal}. The mobility of our devices is an order of magnitude larger than the recently reported mobility ($\sim109cm^2/Vs$) of a wrap-gated device \cite{storm2}, where the fabrication process involved many wet etching steps which can potentially introduce more surface states and thus enhance surface scattering. However it is less than the highest mobility value reported so far \cite{dy,ford} on InAs nanowires. It is known that the reason for reduction of mobility in InAs nanowire from its bulk value is that the surface accumulation charge contributes significantly in electrical conduction, and due to the scattering at the surface the mobility reduces. As a result mobility increases with the wire diameter \cite{ford}. A possible reason for the lower mobility in our case even with a relatively larger diameter of the nanowire ($\sim$70nm) is presumably due to the fact that the twin defect density also increases with the diameter, these twin defects result in additional scattering and a reduction in mobility \cite{peta,thal}. Plot of field effect mobility as a function of temperature, shows that mobility peaks around 50K; a similar trend in the mobility was also observed in a previous work \cite{ford}, and it is attributed to scattering due to surface roughness as well as twin defects.

In conclusion, we have developed a simple fabrication technique for lateral wrap-gate NWFET that can be used for a variety of NW systems. The good capacitive coupling of our devices, and the FET performance including sub-threshold slope and mobility suggest the potential use for high-frequency NWFETs. In addition, such devices with good characteristics at cryogenic temperatures can be used to fabricate on-chip amplifiers for sensitive measurement of current \cite{urazhdin_simple_2002} and capacitance \cite{sulpizio_integrated_2010}. The wrap-gate controlled large electric field will be interesting to study the physics of spin and charge in quasi 1-D systems with spin-orbit interaction  \cite{auslaender_spin-charge_2005,krich_spin-polarized_2008}.

This research work was supported by the Government of India through TIFR grant numbers 11P803, and 11P812. The authors thank Prof. K. L. Narasimhan for discussions about the experiments and the analysis of the results.

\end{document}